\documentclass[preprint,eqsecnum,aps,nofootinbib]{revtex4}
\usepackage{amsfonts,amsmath,amssymb,amsthm}
\usepackage{latexsym}
\usepackage{bbm,bm}
\usepackage{graphicx}

\newcommand{\ket}[1]{\lvert #1 \rangle}
\newcommand{\bra}[1]{\langle #1 \lvert}
\newcommand{\beq}{\begin{equation}}
\newcommand{\eeq}{\end{equation}}
\newcommand{\beqs}{\begin{eqnarray}}
\newcommand{\eeqs}{\end{eqnarray}}

\begin{document}

\title{Tripartite Entanglement in Noninertial Frame}

\author{Mi-Ra Hwang$^1$ and DaeKil Park$^{1,2}$}

\affiliation{$^1$ Department of Physics, Kyungnam University, Masan, 631-701, Korea \\
             $^2$ Department of Electronic Engineering, Kyungnam University, Changwon, 631-701, Korea }

\author{Eylee Jung}

\affiliation{Center for Superfunctional Materials,
Department of Chemistry,
Pohang University of Science and Technology, 
San 31, Hyojadong, Namgu, Pohang 790-784, Korea. 
}

\begin{abstract}
The tripartite entanglement is examined when one of the three parties moves with a uniform acceleration with respect to other 
parties. As Unruh effect indicates, the tripartite entanglement exhibits a decreasing behavior with increasing the acceleration. 
Unlike the bipartite entanglement, however, the tripartite entanglement does not completely vanish in the infinite 
acceleration limit. If the three parties, for example, share the Greenberger-Horne-Zeilinger or W-state initially, the corresponding 
$\pi$-tangle, one of the measures for tripartite entanglement, is shown to be $\pi/6 \sim 0.524$ or $0.176$ in this limit, respectively. 
This fact indicates that the tripartite quantum information processing may be possible even if one of the parties approaches to the Rindler horizon.
The physical implications of this striking result are discussed in the context of black hole physics.
\end{abstract}

\maketitle

\section{Introduction}
It is well known that entanglement of quantum states is a genuine physical resource for various 
quantum information tasks such as quantum teleportation\cite{bennett93}, quantum cryptography\cite{cryptography}, 
and quantum computer technology\cite{qc}. In this reason, recently, much attention is paid to the various properties
of the entanglement\cite{horodecki07}.

In addition to the pure quantum mechanical aspect it is also important to analyze the entanglement of multipartite 
quantum state in the relativistic framework. Evidently, this is an interesting subject from a fundamental point of view. 
Furthermore, it is also important from a practical perspective because many modern experiments on quantum information 
processing use photons or other particles, which have relativistic velocities. The bipartite entanglement between inertial 
frames was investigated in Ref.\cite{inertial}. The remarkable fact in the entanglement between inertial frames is its 
conservation although the entanglement between some degrees of freedom can be transferred to others. Still, however, it is 
not obvious why the entanglement between inertial frames is preserved.  

The bipartite entanglement between noninertial frames was initially studied by Fuentes-Schuller and Mann (FM) in Ref.\cite{schuller04}.
They showed that the maximal bipartite entanglement between inertial parties is degraded if the observers are relatively accelerated.
With increasing acceleration degradation of entanglement becomes larger and larger, and eventually the bipartite state reduces to the 
separable state at infinite acceleration. This phenomenon is sometimes called `Unruh decoherence' and is closely related to the 
Unruh effect\cite{Unruh}. Due to resemblance between Unruh effect and Hawking radiation\cite{hawking} FM predicted that the degradation 
of entanglement occurs in the black hole physics. The degradation phenomenon of bipartite entanglement in a Schwarzschild 
black hole was investigated in Ref.\cite{martinez10}. Although the entanglement is degraded near a Schwarzschild 
black hole as FM predicted, there is a subtle difference arising due to a difference of an event horizon in Schwarzschild spacetime from 
an acceleration horizon in Rindler spacetime. Recently, quantum teleportation between noninertial observers is also discussed 
in detail in Ref.\cite{teleportation}.

In this paper we discuss on the tripartite entanglement in noninertial frame. As far as we know there are two entanglement measures which 
quantify the genuine tripartite entanglement: three-tangle\cite{ckw} and $\pi$-tangle\cite{ou07}. The three-tangle has many nice properties
and is exactly coincides with the modulus of a Cayley's hyperdeterminant\cite{cay1845}. It is also an invariant quantity under the 
local $SL(2,{C})$ transformation\cite{ver03}. Even though its nice features it has a drawback due to its calculational difficulty. 
Since we need an optimal decomposition for analytical computation of the three-tangle for a given tripartite mixed state, it is highly 
difficult problem to compute the three-tangle analytically except few rare cases\cite{rare}. In order to escape this difficulty we adopt the 
$\pi$-tangle for the quantification of the tripartite entanglement solely due to its calculational easiness. The physical roles of the 
three-tangle and $\pi$-tangle in the real quantum information processing was, recently, discussed in Ref.\cite{jung10} in detail.

In this paper we are considering the following situation. Let Alice, Bob and Charlie share  the 
Greenberger-Horne-Zeilinger(GHZ)\cite{green89} or W-state\cite{dur00-1} initially when they are not moving relatively. After then, 
Charlie moves with a uniform acceleration with respect to Alice and Bob. We compute the $\pi$-tangle as a function of Charlie's 
acceleration. It is shown in this paper that the $\pi$-tangle, in general, decreases with increasing the acceleration like the bipartite 
entanglement. However, we show that unlike the bipartite entanglement the $\pi$-tangle does not completely vanish even if Charlie moves 
with an infinite acceleration. This is a striking result in a sense that this fact implies the possibility of the tripartite quantum information processing 
although Charlie approaches to the Rindler horizon. 

This paper is organized as follows.
In section II we consider a situation where Alice, Bob, and Charlie share the GHZ state initially. It is shown that the resulting 
$\pi$-tangle decreases with increasing Charlie's acceleration from $1$ at zero acceleration to $\pi/6 \sim 0.524$ at infinite 
acceleration. In section III the GHZ state in the previous section is replaced with W-state. It is shown that the $\pi$-tangle in this 
case also decreases with increasing acceleration from $4(\sqrt{5} - 1) / 9 \sim 0.55$ at zero acceleration to $0.176$ at infinite acceleration. 
In section VI we discuss the physical implications of the results in the context of the black hole physics.

\section{GHZ state}
In this section we assume that Alice, Bob, and Charlie share initially GHZ state defined as 
\begin{equation}
\label{ghz1}
\ket{GHZ}_{ABC} = \frac{1}{\sqrt{2}} \left[ \ket{000}_{ABC} + \ket{111}_{ABC} \right].
\end{equation}
After sharing his own qubit Charlie moves with respect to Alice and Bob with a uniform acceleration $a$. Then, Charlie's vacuum 
and $1$-particle states $\ket{0}_M$ and $\ket{1}_M$, where the subscript `M' stands for Minkowski, are transformed into\cite{schuller04}
\begin{eqnarray}
\label{unruh1}
& &\ket{0}_M \rightarrow \frac{1}{\cosh r} \sum_{n=0}^{\infty} \tanh^n r \ket{n}_I \ket{n}_{II}      \\   \nonumber
& &\ket{1}_M \rightarrow \frac{1}{\cosh^2 r} \sum_{n=0}^{\infty} \tanh^n r \sqrt{n + 1} \ket{n + 1}_I \ket{n}_{II},
\end{eqnarray}
where $r$ is a parameter proportional to Charlie's acceleration, and $\ket{n}_I$ and $\ket{n}_{II}$ are the mode decomposition 
in the two causally disconnected regions in Rindler space. Eq.(\ref{unruh1}) implies that the physical information formed initially 
in region I is leaked to the inaccessible region (region II) due to accelerating motion. This loss of information causes
a particle detector in the region I to detect a thermally average state, which is a main scenario of Unruh effect\cite{Unruh}.

Therefore, Charlie's acceleration transforms the GHZ state into
\begin{equation}
\label{ghz2}
\ket{GHZ}_{ABC} \rightarrow \frac{1}{\sqrt{2} \cosh r} \sum_{n=0}^{\infty} \tanh^n r
\left[ \ket{00n} \ket{n}_{II} + \frac{\sqrt{n + 1}}{\cosh r} \ket{11n+1} \ket{n}_{II} \right],
\end{equation}
where $\ket{abc} = \ket{ab}_{AB}^M \otimes \ket{c}_I$. Since $\ket{\psi}_{II}$ is a physically inaccessible state from Alice, Bob, and 
Charlie, we should average it out via a partial trace. Thus, the quantum state shared by Alice, Bob, and Charlie reduces to the following 
mixed state:
\begin{eqnarray}
\label{ghz3}
& &\rho_{GHZ} = \frac{1}{2 \cosh^2 r} \sum_{n=0}^{\infty} \tanh^{2n} r
\bigg[ \ket{00n}\bra{00n}                                                        \\    \nonumber
& & \hspace{1.0cm}
+ \frac{\sqrt{n+1}}{\cosh r} \left\{ \ket{00n}\bra{11n+1} + \ket{11n+1}\bra{00n} \right\}
       + \frac{n+1}{\cosh^2 r} \ket{11n+1}\bra{11n+1}             \bigg].
\end{eqnarray}
This is very similar to the information loss of Hawking radiation in the black hole physics, where the pure `in' state becomes thermally
mixed `out' state due to the gravitation collapse\cite{hawking76}.

To quantify how much $\rho_{GHZ}$ is entangled we introduce a $\pi$-tangle\cite{ou07} defined as 
\begin{equation}
\label{pi1}
\pi = \frac{\pi_A + \pi_B + \pi_C}{3}
\end{equation}
where
\begin{eqnarray}
\label{pi2}
& &\pi_A = {\cal N}^2_{A(BC)} - {\cal N}^2_{AB} - {\cal N}^2_{AC}            \\    \nonumber
& &\pi_B = {\cal N}^2_{B(CA)} - {\cal N}^2_{BC} - {\cal N}^2_{BA}            \\    \nonumber
& &\pi_C = {\cal N}^2_{C(AB)} - {\cal N}^2_{CA} - {\cal N}^2_{CB}.            
\end{eqnarray}
In Eq.(\ref{pi2}) ${\cal N}_{\alpha (\beta\gamma)}$ and ${\cal N}_{\alpha\beta}$ are one-tangle and two-tangle respectively, defined as 
${\cal N}_{\alpha (\beta\gamma)} \equiv ||\rho_{GHZ}^{T_{\alpha}} || -1$ and 
${\cal N}_{\alpha\beta} \equiv || (\mbox{tr}_{\gamma} \rho_{GHZ})^{T_{\alpha}} || - 1$. Here $T_{\alpha}$ denotes the partial transposition
for $\alpha$-qubit and $||A||$ is a trace norm of operator $A$ defined as $||A|| \equiv \mbox{tr} \sqrt{A A^{\dagger}}$. 

Although one-tangle can be easily computed in the qubit system by using ${\cal N}^2_{A(BC)} = 4 \mbox{det} \rho^A$, where 
$\rho^A = \mbox{tr}_{BC} \rho_{ABC}$, we cannot use this convenient formula because Charlie's accelerating motion makes the quantum state
infinite-dimensional qudit state. Thus, we have to use the original definition for computation of one-tangle.

Now, let us compute one-tangles. In order to compute ${\cal N}_{A(BC)}$ first we should derive $\rho_{GHZ}^{T_A}$, which is 
\begin{eqnarray}
\label{pghz1}
& &\rho_{GHZ}^{T_A} = \frac{1}{2 \cosh^2 r} \sum_{n=0}^{\infty} \tanh^{2n} r
\bigg[ \ket{00n}\bra{00n}                                                        \\    \nonumber
& & \hspace{1.0cm}
+ \frac{\sqrt{n+1}}{\cosh r} \left\{ \ket{10n}\bra{01n+1} + \ket{01n+1}\bra{10n} \right\}
       + \frac{n+1}{\cosh^2 r} \ket{11n+1}\bra{11n+1}             \bigg].
\end{eqnarray}
From $\rho_{GHZ}^{T_A}$ it is straightforward to derive $\left(\rho_{GHZ}^{T_A}\right) \left(\rho_{GHZ}^{T_A}\right)^{\dagger}$, 
whose matrix representation is a diagonal one. Thus, it is simple to show that the eigenvalues of $\left(\rho_{GHZ}^{T_A}\right) \left(\rho_{GHZ}^{T_A}\right)^{\dagger}$ are
\begin{equation}
\label{ghz-eigen-1}
\left\{ \frac{\tanh^{4n} r}{4 \cosh^4 r}, \frac{(n+1) \tanh^{4n} r}{4 \cosh^6 r}, \frac{(n+1) \tanh^{4n} r}{4 \cosh^6 r},
         \frac{(n+1)^2 \tanh^{4n} r}{4 \cosh^8 r} \bigg| n = 0, 1, 2, \cdots \right\}.
\end{equation}

Using Eq.(\ref{ghz-eigen-1}) one can compute the one-tangle ${\cal N}_{A(BC)}$ by making use of its original definition
$|| \rho_{GHZ}^{T_A}|| - 1$, which is 
\begin{equation}
\label{ghz-one-1}
{\cal N}_{A(BC)} = \frac{1}{\cosh^3 r} \sum_{n=0}^{\infty} \sqrt{n+1} \tanh^{2 n} r.
\end{equation}
When deriving Eq.(\ref{ghz-one-1}) we used the following formulae
\begin{equation}
\label{useful}
\sum_{n=0}^{\infty} \tanh^{2 n} r = \cosh^2 r    \hspace{1.0cm}   
\sum_{n=0}^{\infty} (n + 1) \tanh^{2 n} r = \cosh^4 r.
\end{equation}
Introducing a polylogarithm function $\mbox{Li}_n (z)$ defined as 
\begin{equation}
\label{poly-log}
\mbox{Li}_n (z) \equiv \sum_{k=1}^{\infty} \frac{z^k}{k^n} = \frac{z}{1^n} + \frac{z^2}{2^n} + \frac{z^3}{3^n} + \cdots,
\end{equation}
one can express ${\cal N}_{A(BC)}$ as 
\begin{equation}
\label{ghz-one-2}
{\cal N}_{A(BC)} = \frac{1}{\sinh^2 r \cosh r} \mbox{Li}_{-1/2} (\tanh^2 r).
\end{equation}
By repeating calculation one can show ${\cal N}_{B(AC)} = {\cal N}_{A(BC)}$, which is, in fact, obvious by considering a symmetry 
of the GHZ state.

Now, let us compute the one-tangle ${\cal N}_{C(AB)}$. After deriving $\rho_{GHZ}^{T_C}$ from $\rho_{GHZ}$ given in Eq.(\ref{ghz3}), 
one can construct $\left(\rho_{GHZ}^{T_C}\right) \left(\rho_{GHZ}^{T_C}\right)^{\dagger}$, whose explicit expression is 
\begin{eqnarray}
\label{pghz-2}
& & \left(\rho_{GHZ}^{T_C}\right) \left(\rho_{GHZ}^{T_C}\right)^{\dagger}                     \\    \nonumber
& &= \frac{1}{4 \cosh^4 r} \sum_{n=0}^{\infty} \tanh^{4n} r
\Bigg[ \left(1 + \frac{n \cosh^2 r}{\sinh^4 r} \right) \ket{00n}\bra{00n}                      
+ \left(\frac{n+1}{\cosh^2 r} + \frac{n^2}{\sinh^4 r} \right) \ket{11n}\bra{11n}               \\    \nonumber            
& & \hspace{1.5cm} + \sqrt{n+1} \left(\frac{\sinh^2 r}{\cosh^3 r} + \frac{n}{\cosh r \sinh^2 r} \right) 
\left\{ \ket{00n+1}\bra{11n} + \ket{11n}\bra{00n+1} \right\} \Bigg].
\end{eqnarray}
Unlike the previous cases the matrix representation of $\left(\rho_{GHZ}^{T_C}\right) \left(\rho_{GHZ}^{T_C}\right)^{\dagger}$ is not a 
diagonal matrix. However, one can compute the eigenvalues of $\left(\rho_{GHZ}^{T_C}\right) \left(\rho_{GHZ}^{T_C}\right)^{\dagger}$
analytically by ordering the basis of Hilbert space as
\begin{equation}
\label{basis1}
\left\{ \ket{000}, \ket{110}, \ket{001}, \ket{111}, \ket{002}, \ket{112}, \cdots, \ket{010}, \ket{100}, \ket{011}, \ket{101}, 
\ket{012}, \ket{102}, \cdots \right\}.
\end{equation}
Then, the non-vanishing eigenvalues of  $\left(\rho_{GHZ}^{T_C}\right) \left(\rho_{GHZ}^{T_C}\right)^{\dagger}$ are
\begin{equation}
\label{ghz-eigen-2}
\left\{ \frac{1}{4 \cosh^4 r}, \lambda_n^{\pm} \hspace{.3cm} (n=0, 1, 2, \cdots) \right\}
\end{equation}
where 
\begin{eqnarray}
\label{ghz-eigen-3}
& &\lambda_n^{\pm} = \frac{\tanh^{4n} r}{8 \cosh^4 r}
\Bigg[ \left\{ \frac{2 (n+1)}{\cosh^2 r} + \frac{n^2}{\sinh^4 r} + \tanh^4 r \right\}              \\   \nonumber
& & \hspace{2.0cm} \pm \sqrt{\left\{ \frac{2 (n+2)}{\cosh^2 r} + \frac{n^2}{\sinh^4 r} + \tanh^4 r \right\}
         \left\{ \frac{2 n}{\cosh^2 r} + \frac{n^2}{\sinh^4 r} + \tanh^4 r \right\}      } \Bigg].
\end{eqnarray}  
Thus, the one-tangle ${\cal N}_{C(AB)}$ can be computed straightforwardly from its definition, whose explicit expression is 
\begin{equation}
\label{ghz-one-3}
{\cal N}_{C(AB)} = ||\rho_{GHZ}^{T_C}|| - 1 = 
\frac{1}{2 \cosh^2 r} + \sum_{n=0}^{\infty} \left( \sqrt{\lambda_n^+} + \sqrt{\lambda_n^-} \right) - 1.
\end{equation}
It seems to be impossible to express ${\cal N}_{C(AB)}$ in terms of the polylogarithmic function as the previous cases.

\begin{figure}[ht!]
\begin{center}
\includegraphics[height=10cm]{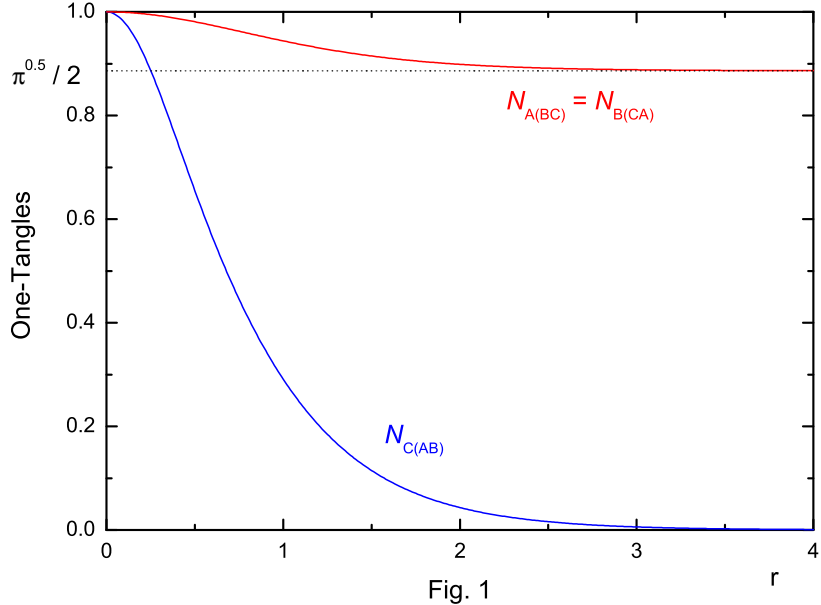}
\caption[fig1]{The $r$-dependence of the one-tangles when Alice, Bob, and Charlie share the GHZ state initially. All one-tangles exhibit
a decreasing behavior with increasing $r$. This figure shows that while ${\cal N}_{C(AB)}$ reduces to zero at $r \rightarrow \infty$ limit, other 
one-tangles do not completely vanish, but goes to $\sqrt{\pi} / 2 \sim 0.886$ in this limit.}
\end{center}
\end{figure}

We plot the $r$-dependence of one-tangles in Fig. 1. All one-tangles become one at $r = 0$, which is exactly the value of 
one-tangles at the rest frame. As expected from the degradation of the bipartite entanglement in noninertial frame all 
one-tangles decrease with increasing Charlie's acceleration. At $r \rightarrow \infty$ ${\cal N}_{C(AB)}$ goes to zero. This 
can be understood from a fact that Alice and Bob cannot contribute to Charlie's entanglement because of Charlie's infinite 
acceleration with respect to Alice and Bob. From this fact we guess that the one-tangle ${\cal N}_{C(AB)}$ goes to zero 
when Charlie falls into a black hole while Alice and Bob are near event horizon of the black hole. This fact also predicts that 
the Coffman-Kundu-Wootters (CKW) inequality\cite{ckw}, ${\cal N}^2_{C(AB)} \geq {\cal N}^2_{CA} + {\cal N}^2_{CB}$, is saturated at 
$r \rightarrow \infty$. As will be shown shortly, this is indeed the case. Surprising fact is that the one-tangles 
${\cal N}_{A(BC)}$ and ${\cal N}_{B(CA)}$ do not vanish but go to $\sqrt{\pi} / 2 \sim 0.886$ in $r \rightarrow \infty$ limit. 
Mathematically, this limiting value is originated from particular properties of the polylogarithmic function. Although we can understand 
that this limiting value is a remnant of entanglement between Alice and Bob without Charlie, we don't know why the remnant is equal to 
this particular value $\sqrt{\pi} / 2$. 

Now, let us compute two-tangles. Since $\rho_{GHZ}^{AB} \equiv \mbox{tr}_C \rho_{GHZ} = (1/2) (\ket{00}\bra{00} + \ket{11}\bra{11})$, 
it is easy to show 
\begin{equation}
\label{ghz-two-1}
{\cal N}_{AB} = || \left( \rho_{GHZ}^{AB} \right)^{T_A} || - 1 = 0.
\end{equation}
Since $\rho_{GHZ}^{AC} = \rho_{GHZ}^{BC}$, ${\cal N}_{AC}$ should be equal to ${\cal N}_{BC}$. One can show that the eigenvalues of 
$\left( \rho_{GHZ}^{AC} \right)^{T_A} \left( \rho_{GHZ}^{AC} \right)^{T_A \dagger}$ are
\begin{equation}
\label{ghz-eigen-4}
\left\{ \frac{\tanh^{4 n} r}{4 \cosh^4 r}, (n+1)^2 \frac{\tanh^{4 n} 4}{4 \cosh^8 r} \bigg| n = 0, 1, 2, \cdots \right\}.
\end{equation}
Using Eq.(\ref{useful}), therefore, one can show easily
\begin{equation}
\label{ghz-two-2}
{\cal N}_{AC} = {\cal N}_{BC} = 0.
\end{equation}
Thus, the two-tangles do not change in spite of Charlie's accelerating motion. 

\begin{figure}[ht!]
\begin{center}
\includegraphics[height=10cm]{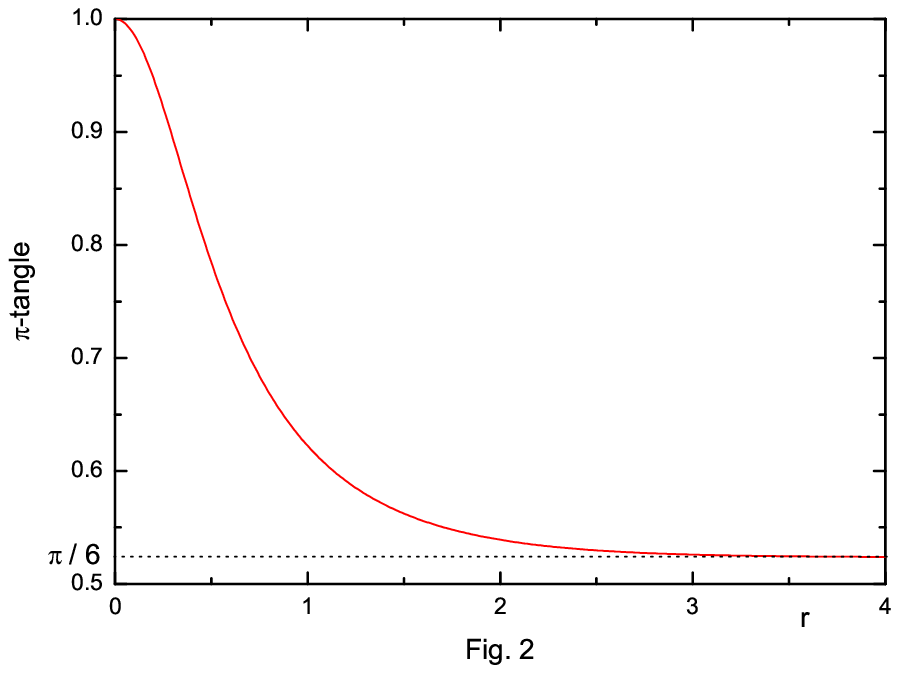}
\caption[fig2]{The $r$-dependence of the $\pi$-tangle when Alice, Bob, and Charlie share the GHZ state initially.
This figure indicates that the $\pi$-tangle does not vanish, but reduces to $\pi/6 \sim 0.524$ in the $r \rightarrow \infty$ limit.
The physical implications of this result are discussed at the final section of this paper.}
\end{center}
\end{figure}

Fig. 2 is a plot of $r$-dependence of $\pi$-tangle when Alice, Bob, and Charlie share initially the GHZ state. As bipartite entanglement the 
$\pi$-tangle decreases with increasing Charlie's acceleration. Unlike bipartite entanglement, however, the $\pi$-tangle does not 
completely vanish in $r \rightarrow \infty$ limit, but approaches to $\pi/6 \sim 0.524$ in this limit. This fact enables us to predict 
that the tripartite entanglement does not completely vanish when Charlie falls into black hole. If so, this is very surprising result 
because this fact implies that quantum communication process might be possible between parties even in the presence of the event horizon. 
This prediction should be checked in the near future by incorporating the quantum information theories in the black hole physics.

\section{W state}
In this section we assume that initially Alice, Bob, and Charlie share W-state
\begin{equation}
\label{w1}
\ket{W}_{ABC} = \frac{1}{\sqrt{3}} \left( \ket{001}_{ABC} + \ket{010}_{ABC} + \ket{100}_{ABC} \right).
\end{equation} 
By making use of Eq.(\ref{unruh1}) one can show that after Charlie's accelerating motion $\ket{W}_{ABC}$ reduces to 
\begin{equation}
\label{w2}
\ket{W}_{ABC} \rightarrow \frac{1}{\sqrt{3} \cosh r} \sum_{n=0}^{\infty} \tanh^n r 
\left[ \frac{\sqrt{n+1}}{\cosh r} \ket{00n+1} + \ket{01n} + \ket{10n} \right] \otimes \ket{n}_{II}
\end{equation}
where $\ket{abc} = \ket{ab}_{AB}^M \otimes \ket{c}_I$. Then, partial trace over $\ket{\psi}_{II}$ transforms W-state into the following
mixed state:
\begin{eqnarray}
\label{w3}
& &\rho_W = \frac{1}{3 \cosh^2 r} \sum_{n=0}^{\infty} \tanh^{2n} r 
\Bigg[ \frac{n+1}{\cosh^2 r} \ket{00n+1}\bra{00n+1} + \ket{01n}\bra{01n} + \ket{10n}\bra{10n}                  \\      \nonumber
& & \hspace{1.5cm} + \frac{\sqrt{n+1}}{\cosh r} \left\{ \ket{00n+1}\bra{01n} + \ket{01n}\bra{00n+1} + \ket{00n+1}\bra{10n} + \ket{10n}\bra{00n+1} \right\}
                                                                                                                \\      \nonumber
& & \hspace{8.0cm} + \left\{ \ket{01n}\bra{10n} + \ket{10n} \bra{01n} \right\}   \Bigg].
\end{eqnarray}
 
Now, let us compute two-tangles. Since $\rho_W^{AB} \equiv \mbox{tr}_C \rho_W$ becomes
\begin{equation}
\label{w-two-1}
\rho_W^{AB} = \frac{1}{3} \left( \ket{00}\bra{00} + \ket{01}\bra{01} + \ket{10}\bra{10} + \ket{01}\bra{10} + \ket{10}\bra{01} \right),
\end{equation}
it is easy to show 
\begin{equation}
\label{w-two-2}
{\cal N}_{AB} = || \left( \rho_W^{AB} \right)^{T_A} || - 1 = \frac{\sqrt{5} - 1}{3}.
\end{equation}
Thus, ${\cal N}_{AB}$ is independent of Charlie's acceleration. In order to compute ${\cal N}_{AC}$ we should derive 
$\rho_W^{AC}$, which can be easily derived from $\rho_W$ by taking a partial trace over Bob's qubit. Then, it is straightforward
to show 
\begin{equation}
\label{w-two-3}
\left( \rho_W^{AC} \right)^{T_A} \left( \rho_W^{AC} \right)^{T_A \dagger}
= \sum_{n=0}^{\infty} \left[ a_n \ket{0n}\bra{0n} + b_n \ket{1n}\bra{1n} + c_n \left\{ \ket{0n}\bra{1n+1} + \ket{1n+1}\bra{0n} \right\}
                                                                                                                                \right]
\end{equation}
where
\begin{eqnarray}
\label{w-two-4}
& &a_n = \frac{\tanh^{4n} r}{9 \cosh^4 r} \left( 1+ \frac{n+1}{\cosh^2 r} + \frac{2n}{\sinh^2 r} + \frac{n^2}{\sinh^4 r} \right)
                                                                                                               \\      \nonumber
& &b_n =  \frac{\tanh^{4n} r}{9 \cosh^4 r} \left( 1 + \frac{n \cosh^2 r}{\sinh^4 r} \right)                    \\      \nonumber
& &c_n = \frac{\sqrt{n+1} \tanh^{4n} r}{9 \cosh^5 r} \left(1 + \tanh^2 r + \frac{n}{\sinh^2 r} \right).
\end{eqnarray}
Although the matrix representation of $\left( \rho_W^{AC} \right)^{T_A} \left( \rho_W^{AC} \right)^{T_A \dagger}$ is not diagonal one, 
one can compute the eigenvalues of it analytically by choosing the order of basis appropriately. Then, the non-vanishing eigenvalues 
of $\left( \rho_W^{AC} \right)^{T_A} \left( \rho_W^{AC} \right)^{T_A \dagger}$ are 
\begin{equation}
\label{w-two-5}
\left\{ b_0, \tilde{\lambda}^{\pm} | n = 0, 1, 2, \cdots \right\}
\end{equation}
where
\begin{equation}
\label{w-two-6}
\tilde{\lambda}^{\pm} = \frac{1}{2} \left[ (a_n + b_{n+1}) \pm \sqrt{(a_n - b_{n+1})^2 + 4 c_n^2} \right].
\end{equation}
Therefore, ${\cal N}_{AC}$ becomes
\begin{equation}
\label{w-two-7}
{\cal N}_{AC} \equiv || \left(\rho_W^{AC} \right)^{T_A} || - 1 = \sqrt{b_0} + \sum_{n=0}^{\infty} 
\left(\sqrt{\tilde{\lambda}_n^+} + \sqrt{\tilde{\lambda}_n^-} \right) - 1.
\end{equation}
Since $\rho_W^{BC} \equiv \mbox{tr}_A \rho_W$ is equal to $\rho_W^{AC}$, it is easy to show ${\cal N}_{BC} = {\cal N}_{AC}$. 

\begin{figure}[ht!]
\begin{center}
\includegraphics[height=10cm]{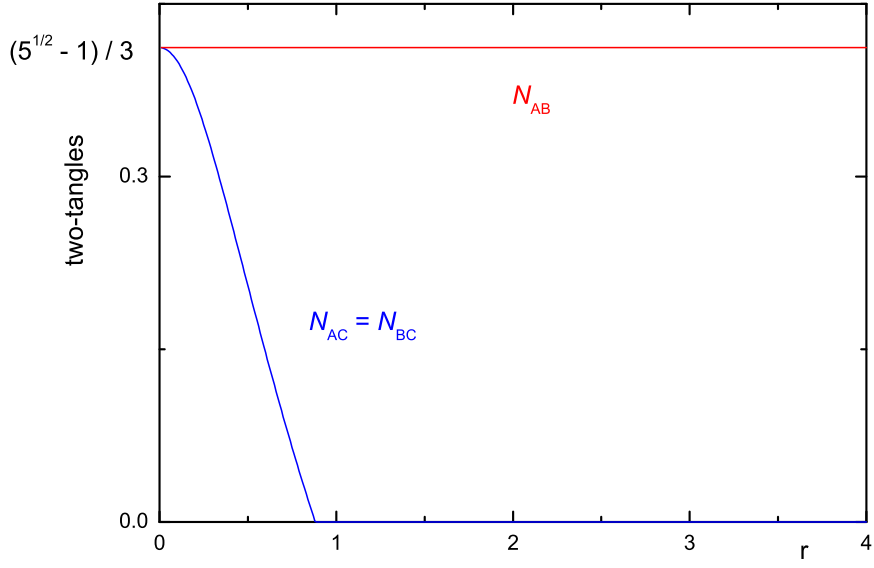}
\caption[fig3]{The $r$-deoendence of the two-tangles when Alice, Bob, and Charlie share the W-state initially.
This figure shows that the two-tangle ${\cal N}_{AB}$ is independent of Charlie's acceleration. However ${\cal N}_{AC}$ and ${\cal N}_{BC}$
decrease with increasing $r$ and become zero at $r \geq 0.89$.}
\end{center}
\end{figure}

The $r$-dependence of the two-tangles is plotted in Fig. 3. When Charlie's acceleration is zero, all two-tangles become
$(\sqrt{5} - 1) / 3 \sim 0.412$, which is two-tangle in the rest frame. As shown in Eq.(\ref{w-two-2}) ${\cal N}_{AB}$ 
is independent of $r$. This is because the contribution of Charlie's qubit average out via the partial trace in $\rho_W^{AB}$. 
However, ${\cal N}_{AC}$ and ${\cal N}_{BC}$ decrease with increasing $r$. This implies that the effect of Charlie's acceleration 
is contributed to these two two-tangles. The remarkable one is the fact that ${\cal N}_{AC}$ and ${\cal N}_{BC}$ become almost zero at 
$r \geq 0.89$. This brings back a concurrence\cite{wootters}, entanglement measure for bipartite quantum state, which is defined as 
$\max (\lambda_1 - \lambda_2 - \lambda_3 -\lambda_4, 0)$, where $\lambda_i$'s are the eigenvalues, in decreasing order, of the 
Hermitian operator $\sqrt{\sqrt{\rho} (\sigma_y \otimes \sigma_y) \rho^* (\sigma_y \otimes \sigma_y) \sqrt{\rho}}$.

Now, let us compute one-tangles. In order to compute ${\cal N}_{A(BC)}$ we should derive $\rho_W^{T_A}$, which can be read 
directly from $\rho_W$ by taking partial transposition for Alice's qubit. Then, after some algebra it is straightforward to show
\begin{eqnarray}
\label{w-three-1}
& &\left(\rho_W^{T_A} \right) \left(\rho_W^{T_A} \right)^{\dagger}                                           \\   \nonumber
& & = \lim_{N \rightarrow \infty} \sum_{n=0}^{N}
\Bigg[ \bar{a_n} \ket{00n}\bra{00n} + \bar{b_n} \ket{01n}\bra{01n} + \bar{c_n} \ket{10n}\bra{10n} + \bar{d_n} \ket{11n}\bra{11n}
                                                                                                              \\   \nonumber
& & \hspace{1.3cm}
+ \bar{f_n} \left\{ \ket{00n+1}\bra{01n} + \ket{01n}\bra{00n+1} \right\} + \bar{g_n} \left\{ \ket{00n}\bra{10n+1} + \ket{10n+1}\bra{00n} \right\}
                                                                                                               \\   \nonumber
& & \hspace{1.3cm}
+ \bar{h_n} \left\{ \ket{00n}\bra{11n} + \ket{11n}\bra{00n} \right\} + \bar{j_n} \left\{ \ket{01n}\bra{10n+2} + \ket{10n+2}\bra{01n} \right\}
                                                                                                                \\   \nonumber
& & \hspace{1.3cm}
+ \bar{k_n} \left\{ \ket{01n}\bra{11n+1} + \ket{11n+1}\bra{01n} \right\} + \bar{\ell_n} \left\{\ket{10n+1}\bra{11n} + \ket{11n}\bra{10n+1} \right\}
\Bigg]
\end{eqnarray}
where
\begin{eqnarray}
\label{w-three-2}
& &\bar{a_n} = \frac{\tanh^{4n} r}{9 \cosh^4 r} \left( 1 + \frac{n+1}{\cosh^2 r} + \frac{n^2 + n \cosh^2 r}{\sinh^4 r} \right)  \hspace{.2cm}
   \bar{b_n} = \frac{\tanh^{4n} r}{9 \cosh^4 r} \left( 1 + \frac{n+1}{\cosh^2 r} \right)                       \\      \nonumber
& &\bar{c_n} = \frac{\tanh^{4n} r}{9 \cosh^4 r} \left( 1 + \frac{n \cosh^2 r}{\sinh^4 r} \right)           \hspace{3.0cm}
   \bar{d_n} = \frac{\tanh^{4n} r}{9 \cosh^4 r}                                                                 \\      \nonumber
& &\bar{f_n} = \frac{\sqrt{n+1} \tanh^{4n} r}{9 \cosh^5 r} \left(1 + \frac{n+1}{\cosh^2 r} \right)         \hspace{.5cm}
   \bar{g_n} = \frac{\sqrt{n+1} \tanh^{4n} r}{9 \cosh^5 r} \left( \tanh^2 r + \frac{n}{\sinh^2 r} \right)        \\      \nonumber
& &\bar{h_n} = \frac{n \tanh^{4n} r}{9 \cosh^4 r \sinh^2 r}                                                 \hspace{2.5cm}
   \bar{j_n} = \frac{\sqrt{(n+1)(n+2)} \tanh^{4n} r}{9 \cosh^8 r} \sinh^2 r                                      \\      \nonumber
& &\bar{k_n} = \frac{\sqrt{n+1} \tanh^{4n} r}{9 \cosh^7 r} \sinh^2 r                                        \hspace{3.0cm}
   \bar{\ell_n} = \frac{\sqrt{n+1} \tanh^{4n} r}{9 \cosh^5 r}.
\end{eqnarray}
It does not seem to be possible to compute the eigenvalues of $\left(\rho_W^{T_A} \right) \left(\rho_W^{T_A} \right)^{\dagger}$ analytically. 
Thus, we adopt a following numerical procedure for the calculation of the eigenvalues. First, we take $N=256$ in Eq.(\ref{w-three-1}) and 
compute numerically $\eta(N, r) = \sum_{i=1}^N \sqrt{\lambda_i} - 1$, where $\lambda_i's$ are the eigenvalues of 
$\left(\rho_W^{T_A} \right) \left(\rho_W^{T_A} \right)^{\dagger}$ and $N=256$. The large $N$-behavior of $\eta(N, r)$ can be computed 
by a numerical fitting  method with using of $\eta(256, r)$. Since ${\cal N}_{A(BC)} = \lim_{N \rightarrow \infty} \eta(N, r)$, the 
$r$-dependence of ${\cal N}_{A(BC)}$ can be computed by following this procedure. The result of the numerical calculation is shown in 
Fig. 4. As Fig. 4 exhibits, ${\cal N}_{A(BC)}$ becomes $2\sqrt{2}/3 \sim 0.943$ at $r=0$. This is a value of one-tangle for W-state in 
the rest frame. As expected it monotonically decreases with increasing $r$, but does not completely vanish at $r \rightarrow \infty$
limit. In this limit ${\cal N}_{A(BC)}$ reduces to ${\cal N}_{A(BC)} \sim 0.659$, which is smaller than $\sqrt{\pi} / 2 \sim 0.886$, 
the corresponding value for GHZ state.

In order to compute ${\cal N}_{B(AC)}$ we should derive $\left(\rho_W^{T_B} \right) \left(\rho_W^{T_B} \right)^{\dagger}$, which can be 
derived straightforwardly from $\rho_W^{T_B}$. The final expression of $\left(\rho_W^{T_B} \right) \left(\rho_W^{T_B} \right)^{\dagger}$
is 
\begin{eqnarray}
\label{w-three-3}
& &\left(\rho_W^{T_B} \right) \left(\rho_W^{T_B} \right)^{\dagger}                                           \\   \nonumber
& & = \lim_{N -> \infty} \sum_{n=0}^{N}
\Bigg[ \bar{a_n} \ket{00n}\bra{00n} + \bar{c_n} \ket{01n}\bra{01n} + \bar{b_n} \ket{10n}\bra{10n} + \bar{d_n} \ket{11n}\bra{11n}
                                                                                                              \\   \nonumber
& & \hspace{1.3cm}
+ \bar{f_n} \left\{ \ket{00n+1}\bra{10n} + \ket{10n}\bra{00n+1} \right\} + \bar{g_n} \left\{ \ket{00n}\bra{01n+1} + \ket{01n+1}\bra{00n} \right\}
                                                                                                               \\   \nonumber
& & \hspace{1.3cm}
+ \bar{h_n} \left\{ \ket{00n}\bra{11n} + \ket{11n}\bra{00n} \right\} + \bar{\ell_n} \left\{ \ket{01n+1}\bra{11n} + \ket{11n}\bra{01n+1} \right\}
                                                                                                                \\   \nonumber
& & \hspace{1.3cm}
+ \bar{j_n} \left\{ \ket{01n+2}\bra{10n} + \ket{10n}\bra{01n+2} \right\} + \bar{k_n} \left\{\ket{10n}\bra{11n+1} + \ket{11n+1}\bra{10n} \right\}
\Bigg]
\end{eqnarray}
where the coefficients are given in Eq.(\ref{w-three-2}). Since $\left(\rho_W^{T_B} \right) \left(\rho_W^{T_B} \right)^{\dagger}$ can be obtained 
from $\left(\rho_W^{T_A} \right) \left(\rho_W^{T_A} \right)^{\dagger}$ by interchanging Alice's qubit and Bob's qubit, the eigenvalues of 
$\left(\rho_W^{T_B} \right) \left(\rho_W^{T_B} \right)^{\dagger}$ should be equal to those of $\left(\rho_W^{T_A} \right) \left(\rho_W^{T_A} \right)^{\dagger}$.
Thus we have ${\cal N}_{B(CA)} = {\cal N}_{A(BC)}$.

\begin{figure}[ht!]
\begin{center}
\includegraphics[height=10cm]{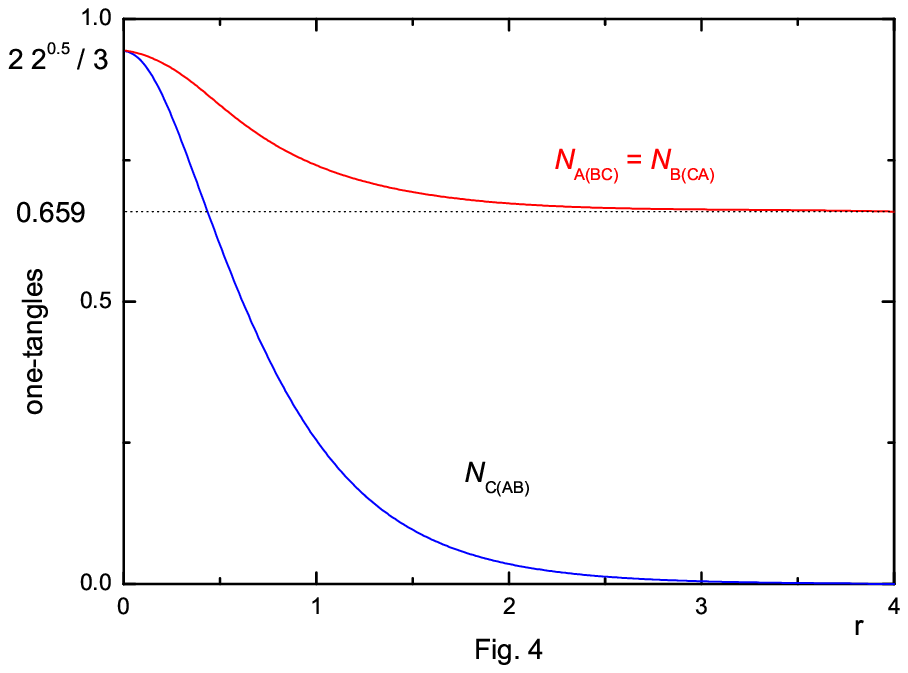}
\caption[fig4]{The $r$-dependence of the one-tangles when Alice, Bob, and Charlie share the W-state initially. Like the case of GHZ state
all one-tangles exhibit a decreasing behavior with increasing $r$. This figure shows that while ${\cal N}_{C(AB)}$ reduces to zero 
at $r \rightarrow \infty$ limit, other one-tangles do not completely vanish, but goes to $0.659$ in this limit.}
\end{center}
\end{figure}

Finally, we compute ${\cal N}_{C(AB)}$. Since $\left(\rho_W^{T_C} \right) \left(\rho_W^{T_C} \right)^{\dagger}$ becomes
\begin{eqnarray}
\label{w-three-4}
& &\left(\rho_W^{T_C} \right) \left(\rho_W^{T_C} \right)^{\dagger}                                              \\    \nonumber
& & =\sum_{n=0}^{\infty} \Bigg[\tilde{a}_n \ket{00n}\bra{00n} + 
\tilde{b}_n \left\{ \ket{01n}\bra{01n}+\ket{10n}\bra{10n}+ \ket{01n}\bra{10n}+\ket{10n}\bra{01n} \right\}        \\    \nonumber
& &\hspace{1.5cm} +\tilde{c}_n \left\{\ket{00n}\bra{01n+1}+\ket{01n+1}\bra{00n}+\ket{00n}\bra{10n+1}+\ket{10n+1}\bra{00n}  \right\}  \Bigg]
\end{eqnarray}
where 
\begin{eqnarray}
\label{w-three-5}
& &\tilde{a}_n = \frac{\tanh^{4n} r}{9 \cosh^4 r} \left( \frac{2(n+1)}{\cosh^2 r} + \frac{n^2}{\sinh^4 r} \right)     \\    \nonumber
& &\tilde{b}_n = \frac{\tanh^{4n} r}{9 \cosh^4 r} \left(2 + \frac{n \cosh^2 r}{\sinh^4 r} \right)                     \\    \nonumber
& &\tilde{c}_n = \frac{\tanh^{4n} r}{9 \cosh^4 r} \left( \frac{2 \sqrt{n+1} \sinh^2 r}{\cosh^3 r} + \frac{n \sqrt{n+1}}{\sinh^2 r \cosh r} \right),
\end{eqnarray}
it is not difficult to compute the eigenvalues of $\left(\rho_W^{T_C} \right) \left(\rho_W^{T_C} \right)^{\dagger}$ analytically by 
choosing the order of the basis appropriately. The final expression of the eigenvalues is 
\begin{equation}
\label{w-three-6}
\left\{ 2 \tilde{b}_0, \Lambda_n^{\pm} \bigg| n=0, 1, 2, \cdots \right\}
\end{equation}
where
\begin{equation}
\label{w-three-7}
\Lambda_n^{\pm} = \frac{1}{2} \left[ \left(\tilde{a}_n + 2 \tilde{b}_{n+1} \right) \pm 
\sqrt{\left(\tilde{a}_n - 2 \tilde{b}_{n+1} \right)^2 + 8 \tilde{c}_n^2} \right].
\end{equation}
Therefore, ${\cal N}_{C(AB)}$ is given by 
\begin{equation}
\label{w-three-8}
{\cal N}_{C(AB)} = \sqrt{2 \tilde{b}_0} + \sum_{n=0}^{\infty} \left( \sqrt{\Lambda_n^+} + \sqrt{\Lambda_n^-} \right) - 1.
\end{equation}

The $r$-dependence of the one-tangles are plotted at Fig. 4. Like GHZ case all one-tangles decrease with increasing $r$. While 
${\cal N}_{C(AB)}$ goes to zero in $r \rightarrow \infty$ limit, ${\cal N}_{A(BC)}$ and ${\cal N}_{B(CA)}$ do not completely vanish but reduce to 
$0.659$ in this limit. This value is smaller than the corresponding remnant $\sqrt{\pi} / 2$ of the one-tangles for the case of GHZ state. As we
commented in the previous section we do not know why the remnant of ${\cal N}_{A(BC)} = {\cal N}_{B(CA)}$ is $0.659$. 

\begin{figure}[ht!]
\begin{center}
\includegraphics[height=10cm]{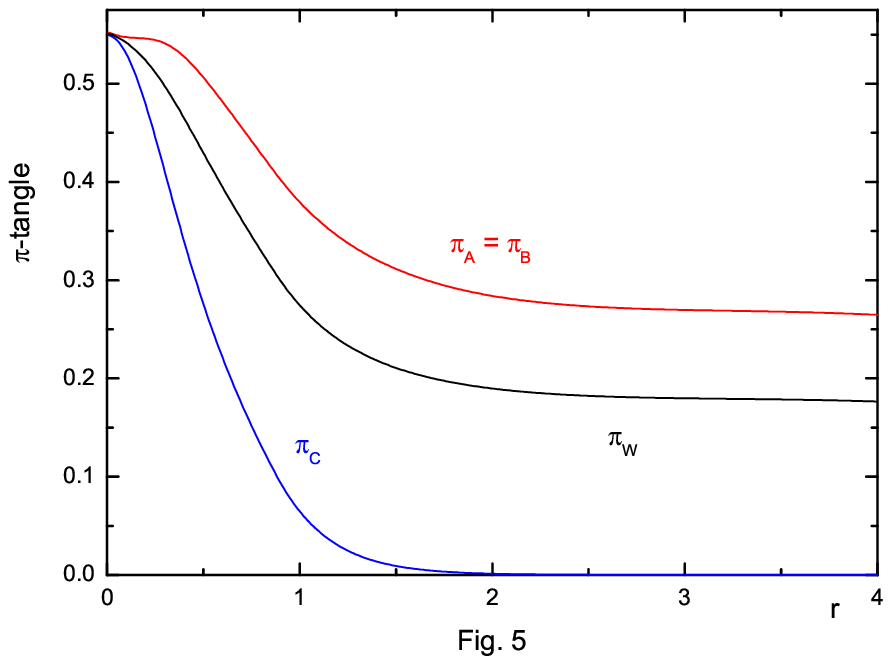}
\caption[fig5]{The $r$-dependence of the $\pi$-tangle when Alice, Bob, and Charlie share the W-state initially. Like the case of GHZ state
the $\pi$-tangle $\pi_W$ exhibits a monotonically decreasing behavior with increasing $r$ and reduces to $0.176$ at $r \rightarrow \infty$ limit.
Mathematically, this is due to the fact that $\pi_A$ and $\pi_B$ become non-zero while $\pi_C$ reduces to zero in this limit. The physical 
implications of this result are discussed in section IV.}
\end{center}
\end{figure}

The $r$-dependence of $\pi$-tangle for W-state is plotted in Fig. 5. 
As expected, $\pi_A$, $\pi_B$, and $\pi_C$ decrease with increasing $r$ from $4 (\sqrt{5} - 1) / 9 \sim 0.55$, which is a 
corresponding value at $r=0$. While $\pi_C$ goes to zero at $r \rightarrow \infty$ limit, $\pi_A$ and $\pi_B$ do not completely vanish 
in this limit, but reduce to $0.265$. In this reason $\pi_W$, the $\pi$-tangle of W-state, becomes $0.176$ when Charlie moves with respect
to Alice and Bob with infinite acceleration. The remnant $0.176$ for W-state is much smaller than the corresponding value $\pi/6 \sim 0.524$
for GHZ state. We do not clearly understand why the remnant of $\pi$-tangle for W-state is much smaller than that for GHZ state. We also do not 
understand why the tripartite entanglement is not zero even when Charlie approaches to the Rindler horizon.


\section{Conclusion}
In this paper we consider the tripartite entanglement when one of the parties moves with uniform acceleration with respect to other parties.
The accelerating motion of the one party is described by Rindler coordinate. We adopt the $\pi$-tangle as a measure of the tripartite entanglement
solely due to its calculational easiness.

Since Unruh effect predicts that the information formed in some region in Rindler space is leaked into the causally disconnected region due to 
acceleration of one party, we expect that the tripartite entanglement decreases with increasing acceleration, and eventually reduces to 
zero at the infinite acceleration limit like the bipartite entanglement\cite{schuller04}. Really, actual calculation reveals the monotonically
decreasing behavior of the $\pi$-tangle with increasing acceleration. However, actual calculation also shows that our expectation of 
vanishment of the $\pi$-tangle in the infinite acceleration limit is wrong. If, for example, the three parties share the GHZ state initially,
the corresponding $\pi$-tangle decreases monotonically from $1$ at zero acceleration to $\pi/6 \sim 0.524$ at infinite acceleration. 
If the parties share W-state initially, the $\pi$-tangle also decreases monotonically from $4(\sqrt{5} - 1) / 9 \sim 0.55$ at zero acceleration 
to $0.176$ at infinite acceleration. Thus, the $\pi$-tangle does not completely vanish even if one of the parties approaches to the Rindler horizon.

The non-vanishment of the $\pi$-tangle at the infinite acceleration is a striking result. Since Rindler spacetime is similar to the 
Schwarzschild spacetime, this result enables us to conjecture that the tripartite entanglement does not completely vanish even if one party
falls into the event horizon of the black hole. If so, some quantum information processing such as tripartite teleportation\cite{tri-teleportation}
can be performed between inside and outside the black hole. Since, however, Rindler horizon is different from the event horizon physically, we 
should check this conjecture explicitly by actual calculation. We would like to re-visit this issue in the near future.

Probably, the non-vanishment of the $\pi$-tangle at the infinite acceleration is due to the incomplete definition of the $\pi$-tangle 
as a measure of the tripartite entanglement. Thus, it seems to be interesting to repeat the calculation of this paper by making use of the 
three-tangle. Since, however, the computation of the three-tangle requires the optimal decomposition of the given mixed state, its 
calculation is much more difficult compared to the $\pi$-tangle. In order to explore this issue, therefore, we should develop analytical
and numerical techniques for the computation of the three-tangle. 

Since recent string and brane-world theories predict the extra dimensions in the spacetime, it seems to be also of interest to study on the 
effect of the extra dimensions in the degradation phenomena of bipartite and tripartite entanglements. Another interesting issue is to explore the 
effect of the black hole's rotation in the bipartite and tripartite entanglements. There are many interesting questions related to this issue. 
For example, it seems to be interesting to examine the relation between superradiance and degradation of entanglement. 
We hope to explore these issues in the future.

\begin{acknowledgments}
This work was supported by National Research Foundation of Korea Grant funded by the
Korean Government (Contract No. F01-2008-000-10039-0).
\end{acknowledgments}

\end{document}